\documentclass[prl,reprint,twocolumn,tightenlines,nopacs,showkeys,superscriptaddress,groupedaddress]{revtex4} 
\usepackage{graphicx, bm} 
\usepackage{dcolumn} 
\usepackage{bm} 
\usepackage{amssymb} 
\usepackage{comment} 
\usepackage[caption=false]{subfig}
\usepackage{amsmath}
\usepackage{float} 
\usepackage{natbib} 
\usepackage{tabularx}
\usepackage{dsfont} 
\usepackage{cancel}

\newcommand{\PRLpar}[1]{{\em #1.}}

\newcommand*\xbar[1]{%
  \hbox{%
    \vbox{%
      \hrule height 0.5pt 
      \kern0.5ex
      \hbox{%
        \kern-0.1em
        \ensuremath{#1}%
        \kern-0.1em
      }%
    }%
  }%
} 

\begin{document}

\title{Universal properties of concentration sensing in large ligand-receptor networks}

 \author{Vijay Singh} \affiliation{Department of Physics, \& Computational Neuroscience Initiative, University of
  Pennsylvania, Philadelphia, PA 19104, USA}
\email{vsin@sas.upenn.edu}
\author{Ilya  Nemenman} \affiliation{Department of Physics, Department of Biology, and Initiative in Theory and Modeling of Living Systems, Emory University, Atlanta, GA 30322, USA}
\email{ilya.nemenman@emory.edu}

\begin{abstract}
Cells estimate concentrations of chemical ligands in their environment using a limited set of receptors. Recent work has shown that the temporal sequence of binding and unbinding events on just a single receptor can be used to estimate the concentrations of multiple ligands. Here, for a network of many ligands and many receptors, we show that such temporal sequences can be used to estimate the concentration of a few times as many ligand species as there are receptors. Crucially, we show that the spectrum of the inverse covariance matrix of these estimates has several universal properties, which we trace to properties of Vandermonde matrices. We argue that this can be used by cells in realistic biochemical decoding networks.
\end{abstract}
\keywords{cellular information processing, random matrices, maximum likelihood, Vandermonde matrices}

\maketitle

\PRLpar{Introduction} 
Cellular environment contains many chemical ligands that are  sensed by cell surface receptors. Typically, the number of ligand species is larger than that of the receptors. However, in traditional treatments of the problem, one only takes into account the (fluctuating) steady-state occupancy of  receptors \cite{Berg:1977bp, Bialek2005, hu2010physical, kaizu2014berg}, which allows  estimation of just one quantity (e.~g., one ligand species) per receptor. Recent work has focused instead on using sequences of binding and unbinding times for the estimation. The durations of unbound times carry information about the  concentrations \cite{endres2009maximum}, while the durations of bound times identify the ligands. This allows to estimate the concentrations even in the presence of spurious background ligands \cite{Lalanne:2015ut, Mora:2015cv}, or to get information about more than one concentration simultaneously from a single receptor \cite{singh2017simple,CarballoPacheco:2018vi}. All of this can be done using biologically plausible chemical reaction networks, such as variations of the kinetic proofreading model \cite{hopfield1974kinetic,ninio1975kinetic}.

Here we turn to a previously not investigated regime, where both the number of ligands and the number of receptors are large, and ligands and receptors interact with a broad distribution of binding affinities, the so called Multiple Inputs -- Multiple Outputs (MIMO) problem. We use maximum-likelihood (ML) techniques \cite{endres2009maximum,Mora:2015cv,singh2017simple} to estimate concentrations of all ligands from outputs of all receptors, which provides estimates that are consistent with the true concentrations \cite{wald1949note}. We then focus on the co-variance of the estimates, obtained from the Hessian matrix. Our main finding is that the eigenvalue spectrum of this matrix exhibits universal behaviors, which we trace to properties of Vandermonde matrices. We argue that such MIMO problem is common in various biological systems, and it is also of relevance more broadly, beyond the ligand-receptor problem we study here.  

\PRLpar{Model} 
Consider a mixture of $N_{\rm L}$ ligands, with concentrations $c_{\rm \alpha}, {\rm \alpha}\in[1,N_{\rm L}]$, 
that bind to $N_{\rm R}$ receptors with binding (unbinding) rates $k_{\alpha i}\;(r_{\alpha i})$. Notice that $N_{\rm L}$ can be larger than $N_{\rm R}$, so that more ligand concentrations are being measured than there are receptor types. Further, ligands and receptors cross-react, so that, in principle, $k_{\alpha i }>0$ and  $r_{\alpha i}<\infty$ for all $\alpha$ and $i$. Thus we do not distinguish cognate and noncognate ligand-receptor pairs.

Suppose $n_i$ binding/unbinding events happen on the $i$'th receptor over the measurement time $T$. There is no way of knowing which particular ligand caused which binding, but binding durations are known, which provides some information about the ligand identity. In the regime of interest  $n_i\gg 1$ and a single binding or unbinding event matters little. Thus  we assume that all receptors are unbound at $t=0$ and bound at $t=T$. Then the sequence of unbound/bound durations of the $i$'th receptor is
$\{\vec{\tau}_i^{\rm u},\vec{\tau}_i^{\rm b}\}=\{\tau^{\rm u}_{1_i},\tau^{\rm b}_{1_i}, \tau^{\rm u}_{2_i},\tau^{\rm b}_{2_i},\dots ,\tau^{\rm u}_{n_i},\tau^{\rm b}_{n_i}\}$.
The likelihood of observing such a sequence is \cite{singh2017simple}:
\begin{align}
P(\{\vec{\tau}_i^{\rm u},\vec{\tau}_i^{\rm b}\}|\{c_{\rm{\alpha}}\})&=\frac{1}{Z}\prod_{i = 1}^{N_{\rm R}} \prod_{m_i=1}^{n_i} 
\left[e^{-\tau^{\rm u}_{m_i} \sum\limits_{\alpha=1}^{ N_{\rm L}} c_{\alpha}  k_{\alpha i}} \right. \nonumber \\
         & \left. \times  \sum\limits_{\alpha=1}^{N_{\rm L}} c_{\alpha}  k_{\alpha i} r_{\alpha i} e^{-\tau^{\rm b}_{m_i}r_{\alpha i}} \right].
\label{eq:prob}
\end{align}

 \begin{figure*}
\centering
\includegraphics[width=1\textwidth, trim={0cm 2.0cm 0 0}]{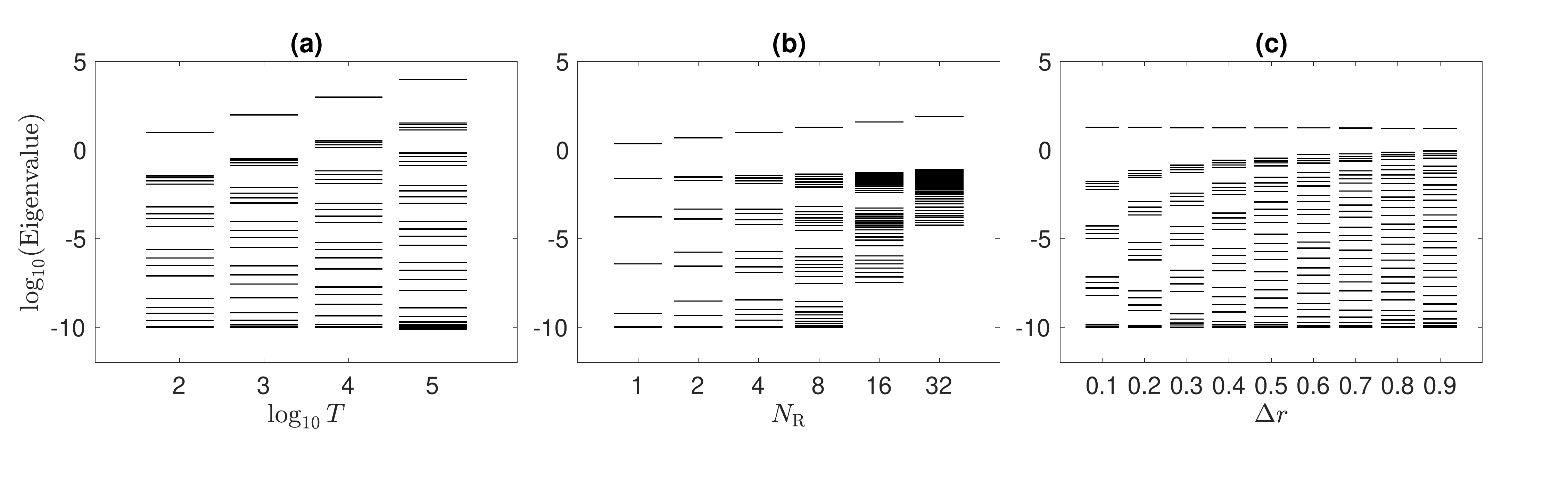}%
\caption{(a) Eigenvalue spectrum of the Hessian matrix $\log P''$ plotted for different total observation time $T$ for a network of 4 receptors and 40 ligands. The eigenvalues are averaged over 10 random realizations of the unbinding rates. All concentrations and binding rates are set to 1. The unbinding rates are chosen from a log-normal distribution with mean parameter 1 and the standard deviation parameter 0.1. In these simulations, we have set the minimum eigenvalue (the inverse of the prior variance) as $10^{-10}$. In reality, eigenvalues much smaller than 1 will be dominated by the prior and are not physically relevant.  The largest eigenvalue corresponds to measuring the total concentration of all ligands. Other eigenvalues group together in subsets of the number of receptors (here $N_{\rm R} = 4$). These subsets are nearly equally spaced on the log axis. (b) Averaged (over 1000 random realizations of the unbinding rates)  eigenvalue spectrum of $\log P''$ vs the number of receptors, $N_{\rm R}$. Here $N_{\rm L}=40$ and $T=100$. The concentrations and the unbinding rates are as above. As the number of receptors changes, the size of the split subsets follows. (c) Averaged (over 1000 random realizations of the unbinding rates) eigenvalue spectrum of $\log P''$ vs the standard deviation of the unbinding rate distribution, which is log-normal with the mean parameter 1. We simulated a network of 4 receptors and 40 ligands for $T=100$. The concentrations and binding rates were chosen as earlier.  A network with wider range of unbinding rates estimates concentrations better (larger eigenvalues). \label{fig:networkSimulation}
}
 \end{figure*}

The log likelihood can be maximized to get the maximum a posteriori (MAP) estimate of the concentration vector $\{c_{\alpha}\}$. Differentiating Eq.~(\ref{eq:prob}) w.~r.~t.\ $c_\alpha$ gives the following $N_{\rm L}$ coupled algebraic MAP equations:
\begin{multline}
0=\frac{\partial \log P}{\partial c_\alpha}\equiv (\log P)'_{\alpha}= \frac{\partial {\log\mathcal{P}}}{\partial c_{\alpha}} +\\\quad 
\sum_{i=1}^{N_{\rm R}} \sum_{m_i=1}^{n_i} \left(
  -\tau^{\rm u}_{m_i} k_{\alpha i}  + \frac{k_{\alpha i} r_{\alpha i} e^{-\tau^{\rm b}_{m_i}r_{\alpha i}}}{
  \sum_{\alpha=1}^{ N_{\rm L}} c_{\alpha}  k_{\alpha i} r_{\alpha i} e^{-\tau^{\rm b}_{m_i}r_{\alpha i}}} \right),
\label{eq:MLEqs}
\end{multline}
where $\mathcal{P}$ is the prior over the concentrations.
The co-variance of the estimation can be obtained from the inverse of the Hessian matrix $(\log P'')_{\alpha\beta}$ evaluated at the MAP solution.
We get:
\begin{equation}
\log P'' = \sum_{i=1}^{N_{\rm R}} \log P''_{i} + \log \mathcal{P}'',
\label{eq:matrixSum}
\end{equation}
where $\left(\log \mathcal{P}''\right)_{\alpha\beta} = \frac{\partial^2\log\mathcal{P}}{\partial c_{\alpha}\partial c_{\beta}} \delta_{\alpha\beta}$
and
\begin{equation}
\left(\log P''_{i}\right)_{\alpha\beta} = \sum_{m_i=1}^{n_i} \frac{(k_{\alpha i} r_{\alpha i} e^{-\tau^{\rm b}_{m_i}r_{\alpha i}})(k_{\beta i} r_{\beta i} e^{-\tau^{\rm b}_{m_i}r_{\beta i}})}{
  \left( \sum_{\alpha = 1}^{N_{\rm L}} c_{\alpha}  k_{ \alpha i} r_{\alpha i} e^{-\tau^{\rm b}_{m_i}r_{\alpha i}} \right)^2}.
\label{eq:hessian}
\end{equation}
We assumed that the {\em a priori} concentration co-variances are zero.
$\log P''_i$ contains terms that are contributed to the $\alpha\beta$ entry in the Hessian matrix by the receptor $i$. 
The sum in Eq.~(\ref{eq:hessian}) is over the durations for the which this $i$'th receptor was bound, because only the bound durations are informative of the ligand identity. 

Variances of the estimates scale as the inverse of the eigenvalues of $\log P''$ -- high eigenvalues correspond to lower variance of various linear combination of the estimated concentrations. Thus we focus on the spectrum of eigenvalues of $\log P''$ in the rest of the paper.

\begin{figure}[t]
\centering
\includegraphics[width=0.35\textwidth, trim={0cm 0.5cm 0 0}]{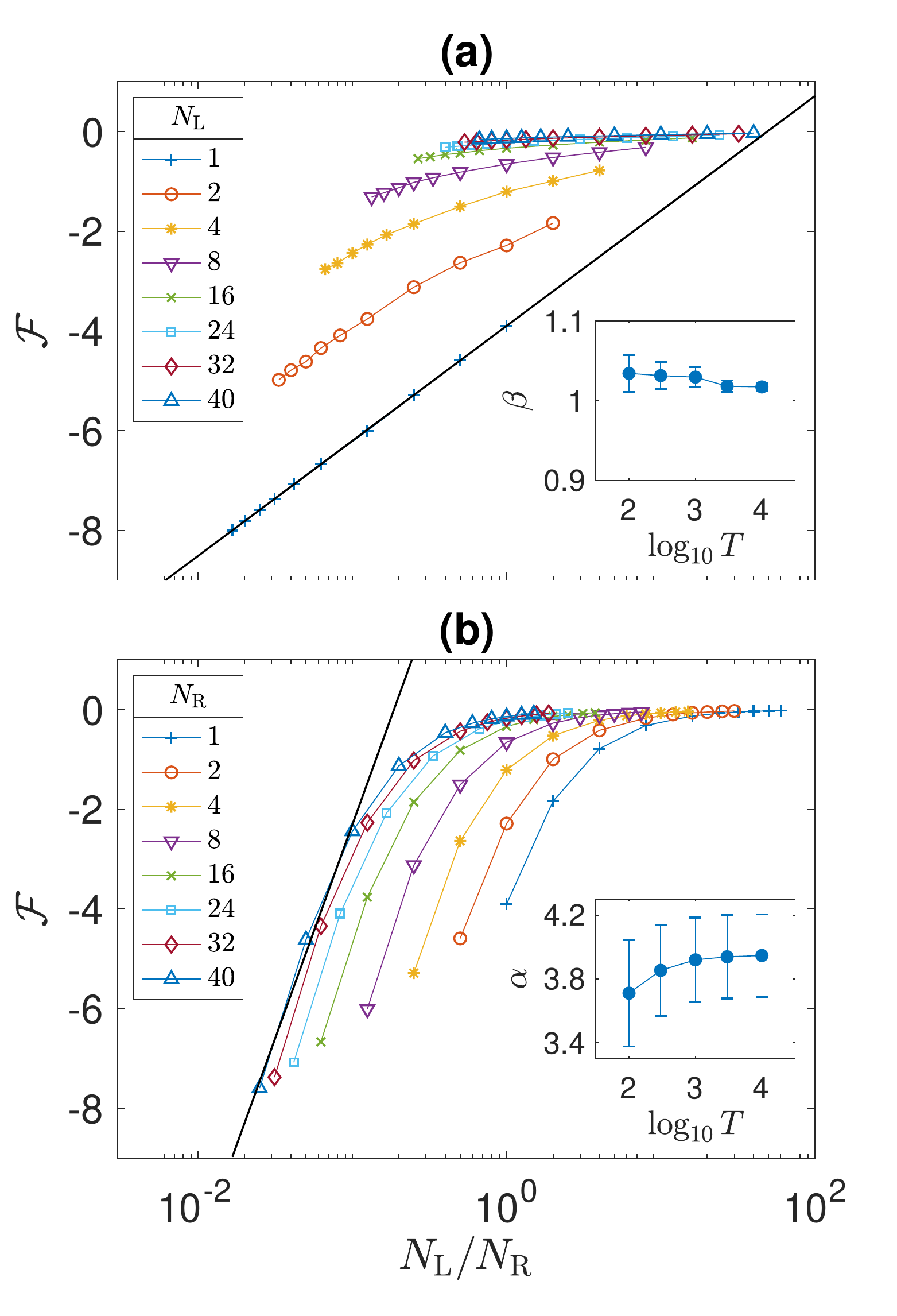}%
   \caption{Asymptotic properties of the inference are reflected in the negative logarithm of the determinant of the Fisher information matrix per ligand species, ${\mathcal F}$. We plot ${\mathcal F}$ vs.\  $N_{\rm L}/N_{\rm R}$, connecting the data points with (a) constant $N_{\rm L}$ and (b) constant $N_{\rm R}$. Parameters of the simulations are as in Fig.~\ref{fig:networkSimulation}, with the log-variance of the log-normal distribution of the unbinding rate $\Delta r = 0.1$, and $T=100$. Asymptotically, ${\mathcal F}$ scales as $\left(N_{\rm L}\right)^{\alpha}$, $\alpha\approx 3.7$, and as $1/\left(N_{\rm R}\right)^{\beta}$, $\beta \approx 1$. Insets in both panels show how $\alpha$ and $\beta$ change with $T$. Error bars represents ($\pm $) 1 standard error obtained from the linear fits.\label{fig:DetVsL}}
\end{figure}

\PRLpar{Eigenvalue spectrum of the inverse covariance matrix}
To illustrate the main properties of the eigenvalues of $\log P''$, we performed simulations of ligand-receptor networks of varying sizes. 
We explored different distributions of binding rates, concentrations, and unbinding rates, and these have little effect on the conclusions drawn below, unless noted otherwise. 
First, we set all (nominally unknown) ligand concentrations and binding rates to 1 for simplicity. This is the hardest inference problem -- the least certainty about which specific ligand got bound to which specific receptor. 
Further, we choose to work with the log-normal distribution of unbinding rates (or, equivalently, normally distributed energy barrier between bound and unbound states). 
Figure~\ref{fig:networkSimulation} shows the eigenvalues of the matrix $\log P''$ vs the total simulation time $T$ for a network of 40 ligands and 4 receptors in panels (a) and (c), and a variable number of receptors in panel (b). 
In all cases, the highest singleton eigenvalue for each $T$ corresponds to the estimate of the total concentration, $c_{\rm tot}=\sum c_\alpha$. However, the rest of the eigenvalues come in subsets, whose size is equal to the number of receptors in the network, and these sets are separated almost equidistantly on a log-scale. Further, the mean of the eigenvalues and the top singleton eigenvalue are higher for systems with more receptors -- which corresponds to a better inference coming from more independent samples of the concentrations, cf.~Fig.~\ref{fig:DetVsL}. Finally, as the variance of the unbinding rates increases, the eigenvalues get lifted (the overall inference improves), and the splitting of subsets decreases, which suggests that (i) diversity of cross-reactivities among ligands and receptors improves the estimation, and (ii) the splitting is due to the degeneracy of nearly similar unbinding rates.

We now assume that the concentrations $c_\alpha$ are sampled from an unknown Gaussian distribution, $\mathcal{N}(\mu=1,\sigma=1)$. We can quantify the ability to infer the concentrations from the binding data by calculating the Fisher information -- the expected value (over the concentration prior) of the Hessian of the log-likelihood matrix at the MAP solution. Specifically, we focus on the negative logarithm of the  determinant of the Fisher information matrix, divided by the number of ligands, ${\mathcal F} \equiv  -\langle\log P''_{\alpha\beta}|_{\rm MAP}\rangle/N_{\rm L}$. It represents  the average logarithm of the variance on the ligand concentration following an observation. We investigate this quantity as a function of $N_{\rm R}$ and $N_{\rm L}$ in Figure~\ref{fig:DetVsL}. We observe that ${\mathcal F}\propto {N_{\rm L}}^{\alpha}/N^\beta_{\rm R}$, with $\alpha \approx 3.7$ and $\beta\approx 1$ for the observation duration $T=100$. The dependence on $N_{\rm R}$ represents the usual law of large numbers -- as more measurements of the concentrations are made, the average variance of the estimates, ${\mathcal F}$, falls inversely proportionally with the number of measurements, $N_{\rm R}$. Correspondingly, $\beta$ does not change with $T$. The reason for scaling with $N_{\rm L}$ is unclear, but it indicates that decreasing the number of ligands has a dramatic effect on the ability to estimate them, approaching $\alpha\approx 4$ at asymptotically large $T$.   Notice also that ${\mathcal F}$ starts deviating from 0 at $N_{\rm L}/N_{\rm R} \sim 3\dots10$, indicating that the network can estimate nearly $3\dots 10$ as many  ligand species as it has the receptors.
Qualitatively similar results are obtained for other types  distributions of the unbinding rates \cite{supplement}.

\begin{figure}[t]
\centering
\includegraphics[width=0.45\textwidth, trim={0 0cm 0 0}]{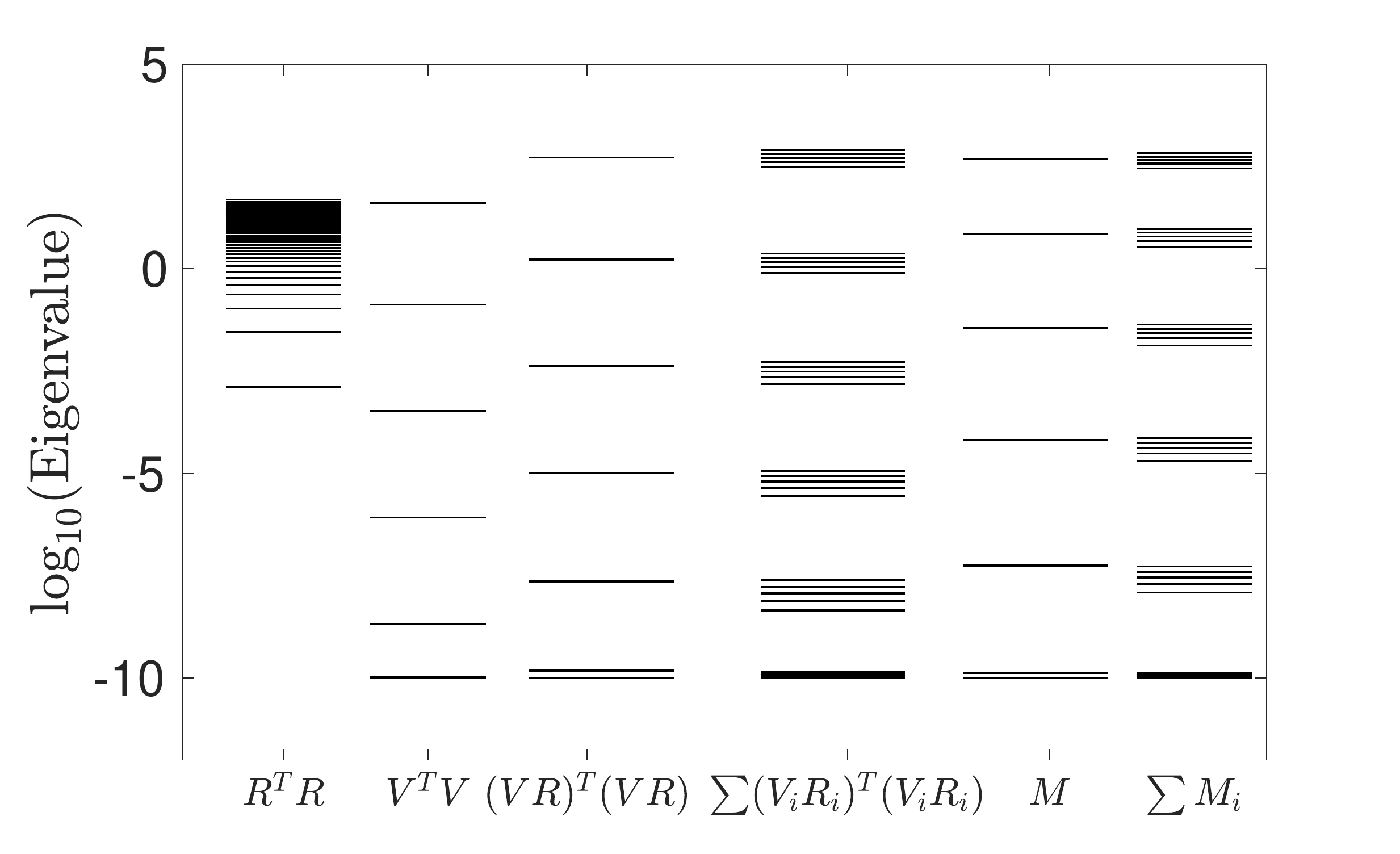}
   \caption{ The eigenvalue structure observed for the Hessian for the ligand-receptor networks has its origin in the Vandermode matrix, through the expansion of the type $(VR)^T(VR)$ (see text).  The first four columns show spectra of $R^TR$, where $R$ is a random matrix, $V^TV$ \cite{waterfall2006sloppy}, $(VR)^T(VR)$, and a sum of different $(VR)^T(VR)$ matrices. $R$ and $V$ are $40 \times 40$ square matrices and 5 such matrices are added for the spectra in the fourth column. Such matrices are ill conditioned due to the $V^TV$ factor, and their eigenvalue spectra are dominated by this factor. Notice that the spectrum of the $V^TV$ factor determines the spectrum of  $(VR)^T(VR)$, and adding many such matrices results in level splitting. A related argument is illustrated in the last two columns. We suggest that the eigenvalues of $\log P''_i$ result from the exponential part in the numerator of Eq.~(\ref{eq:hessian}). Matrix $M_{40 \times 40}$ in fifth column has the same structure $(M_i)_{\alpha\beta} = \sum_{m=1}^{N} (x_{\alpha}x_{\beta})^{\tau_m}$, resulting in exponentially spaces eigenvalues. For these simulations, we chose $\tau$ uniformly at random in [0 1], $x$ uniformly at random in [0.9, 1.1], and $N=20$. Adding many such matrices together again results in level splitting (last column).}
   \label{fig:EigValRV}
 \end{figure}

\PRLpar{Origin of the eigenvalue spectrum}  Two related arguments can explain our observations that (i) eigenvalues are split in groups of size $N_{\rm R}$, (ii) the groups are nearly equidistant from each other on the log space, (iii) higher diversity of unbinding rates decreases the gap between the groups. We start by writing 
$\log P''_i = J_i^T J_i$, where
\begin{equation}
(J_i)_{m_i\alpha} = \frac{ k_{\alpha i} r_{\alpha i} e^{-\tau^{\rm b}_{m_i}r_{\alpha i}}}
{ \sum_{\gamma = 1}^{N_l} c_{\gamma}  r_{\gamma i} e^{-\tau^{\rm b}_{m_i}r_{\gamma i}}}.
\end{equation}
Assuming that the bound time durations on the receptor $i$, $\tau^{\rm b}_{m_i}$  are 
distributed narrowly around some mean value $\bar{\tau}^{\rm b}_i$ (the accuracy of this assumption depends on the tightness of the distribution of the unbinding rates), 
we can expand $(J_i)_{m_i\alpha }$ around this mean value. This gives
\begin{equation}
(J_i)_{m_i\alpha}  =  \sum \limits_{\nu =1}^{\infty} (J_i)^{(\nu-1)}_{\alpha} (d\tau_{m_i})^{\nu-1},
\label{eq:JExpansion}
\end{equation}
where $(J_i)^{(\nu)}_{\alpha}$ represents the $\nu$'th derivative of $(J_i)_{m_i\alpha}$, evaluated at $\bar{\tau}^{\rm b}_i$,  and $d\tau_{m_i}= \tau_{m_i}^{\rm b}-\bar{\tau}^{\rm b}_i$,

Using the expansion in Eq.~(\ref{eq:JExpansion}), we can write  $J_i = V_iA_i$, where $\left(A_i\right)_{\nu\alpha}=(J_i)^{(\nu-1)}_{\alpha}$ and $V_i$ is the Vandermonde matrix 
$(V_i)_{m_i\nu}=\left(d\tau_{m_i}\right)^{\nu-1}$ \cite{lam1985general}. 
So,  $\log P''_i = J_i^T J_i = (A_i)^T(V_i)^TV_iA_i$. 
The eigenvalues of the matrix $(V_i)^TV_i$ scale as $d\tau^{2(\nu-1)}$ \cite{waterfall2006sloppy}. 
We performed simulations with matrices of this form, namely $(VR)^T(VR)$, where $R$ is a random matrix with elements chosen uniformly at random from $[0, 1]$, and $d\tau_i$ in the Vandermonde matrices were chosen uniformly at random from $[-0.1, 0.1]$. Other ranges and distributions of values produce qualitatively similar results.
The resulting eigenvalues 
are shown in Fig.~\ref{fig:EigValRV}, together with  eigenvalues of matrices $V$ and $R$, for comparison. 
These simulations suggest that the eigenvalues of $\log P''_i$ follow the same scaling as of $V^TV$.

The matrix formed by adding several matrices of the form $\log P''_i$ has an eigenvalue spectrum similar to that of $\log P''$ (compare Fig.~\ref{fig:EigValRV} to Fig.~\ref{fig:networkSimulation}). 
This is because the matrices being added ($\log P''_i$) have eigenvalues distributed roughly exponentially, but the corresponding eigenvectors are rotated randomly w.~r.~t.~each other. In high-dimension, such random rotations result in the eigenvectors  corresponding to the eigenvalue of the same rank being almost orthogonal to each other. This will introduce level splitting, similar to degenerate perturbation theory in quantum mechanics, so that if $N$ such matrices are added, eigenvalues will come in groups of $N$ sets. 

We can illustrate the same result with a different, but related argument. We simulated matrices of the form 
$(M_i)_{\alpha\beta} = \sum_{m=1}^{N} (x_{\alpha}x_{\beta})^{\tau_m}$, 
where $x_{\alpha}$'s and $\tau$'s are generated randomly, and $N\gg1$ (our conclusions below  hold for $N\gtrsim5$). This form corresponds to the exponential terms in  Eq.~(\ref{eq:hessian}). The eigenvalues of these simulated matrices show an exponential scaling similar to that of $\log P''_i$, cf.~Fig.~\ref{fig:EigValRV}. 
Further, if several such matrices are added together, $M = \sum_{i}M_{i}$, the resulting eigenvalue spectrum again looks similar to that of $\log P''$ because of the same level splitting argument for orthogonal eigenvectors corresponding to the same eigenvalues. This suggest that the eigenvalue spectrum we see for the ligand-receptor network is not overly specific to this system, but results from the particular structure (sum of exponentiated bilinear terms) of the Hessian matrices.

\PRLpar{Decoding concentrations from receptor activities} 
Our receptor-ligand MIMO inference scheme only measures the duration of time for which each receptor is bound and unbound. Moments of these times can be used to infer concentrations of individual ligand species:
\begin{multline}
\langle(\tau_{i}^{\rm b})^n\rangle = \sum_{\alpha}  \frac{k_{\alpha i} c_{\alpha}}{\sum_{\alpha'} k_{\alpha' i} c_{\alpha'}} \int \limits_{0}^{\infty} \tau_{m_i}^n r_{\alpha i} e^{-r_{\alpha i} \tau_{m_i}}d\tau_{m_i} \\ = 
\sum_{\alpha} \frac{c_{\alpha}}{c_{\rm tot}} \frac{1}{r_{\alpha i}^n} \int \limits_{0}^{\infty} \tau^n e^{-\tau}d\tau = \sum_{\alpha} \frac{c_{\alpha}}{c_{\rm tot}} \frac{\Gamma(n+1)}{r_{\alpha i}^n},
\label{eq:moments}
\end{multline}
where the term outside the integral in the first line is the probability of ligand $\alpha$ to bind receptor $i$ and
$r_{\alpha i} e^{-r_{\alpha i} \tau_{m_i}}$ is the probability density for such bound interval durations.
Here we assumed the binding rates to be the same for simplicity, which happens when they are diffusion-limited. Thus each moment is a linear combination of the ligand concentrations.
 
It is easy to design realistic biochemical networks that would solve such a systems of linear equations and infer the concentrations \cite{Mora:2015cv,singh2017simple,CarballoPacheco:2018vi}.
For example, to estimate the first moment $\langle \tau_i\rangle$, a reporter molecule can be generated only when the receptor $i$ is bound. The mean reporter amount produced over time $T$ would be proportional to $T$, and, assuming many such molecules are produced over a typical bound interval, noise due to discreteness of the reporter would be negligible. Similarly, the estimate of the second moment, $\langle \tau_i^2\rangle$ can be obtained from a secondary reporter, which gets produced with the rate proportional to the instantaneous amount of the first reporter, and only while the receptor is bound. If the production rates are high, discreteness of this reporter will also be negligible. Then final network readouts can be activated / suppressed by the reporters to form their appropriate linear combinations representing $c_\alpha$\cite{singh2017simple}.

\PRLpar{Discussion}
Here we studied a network of ligands and receptors with crosstalk, such that the number of ligands is larger than the number of receptors. Using the maximum likelihood solution based on time series of receptor binding and unbinding, we showed that estimation of ligands in this context, which could have been underdetermined, is, in fact, possible.  We noticed that the maximum likelihood estimator results in a Hessian with a spectrum with a universal properties. Specifically, the eigenvalues come in subsets whose size is equal to the number of receptors in the network, and these subsets are almost exponentially distributed. This observation can be employed by biological systems to design simple and universal chemical kinetics  schemes that would estimate concentrations of ligands from receptor activities irrespectively of the details of binding and unbinding rates in the network. Thus our theory makes specific predictions about the structure of molecular networks downstream of cell surface receptors. Whether biological systems follow these predictions remains to be seen, and networks around EGFR, IGF-1R, FC, and BMPR families of receptors \cite{EGFR,IGF1R,FC,BMPR}, which respond to many related ligands by many related receptor, are good targets for the investigation. Finally, since we traced the universal spectrum of the Hessian to the properties of Vandermonde matrices that enter the inference problem, we expect that our findings will be applicable more generally, beyond ligand-receptor molecular networks, including problems in neuroscience and artificial sensory networks.

\PRLpar{Acknowledgements} We thank Martin Tchernookov, Aditya Rangan, Thierry Mora and William S.\ Hlavacek for valuable discussions. VS was supported by University of Pennsylvania CNI fellowship. IN was supported in part by NIH and NSF Grants 1R01-EB022872, 5R01-NS099375, and PHY-1410978. IN acknowledges hospitality of the Kavli Institute for Theoretical Physics, supported in part by NSF and NIH Grants PHY-1748958, R25GM067110, and the Gordon and Betty Moore Foundation Grant 2919.01.

\bibliography{manyReceptors.bib}

\clearpage

\renewcommand{\thefigure}{A\arabic{figure}}
\setcounter{figure}{0}

\begin{figure*}
\centering
\includegraphics[width=\textwidth, trim={0 2.0cm 0 0}]{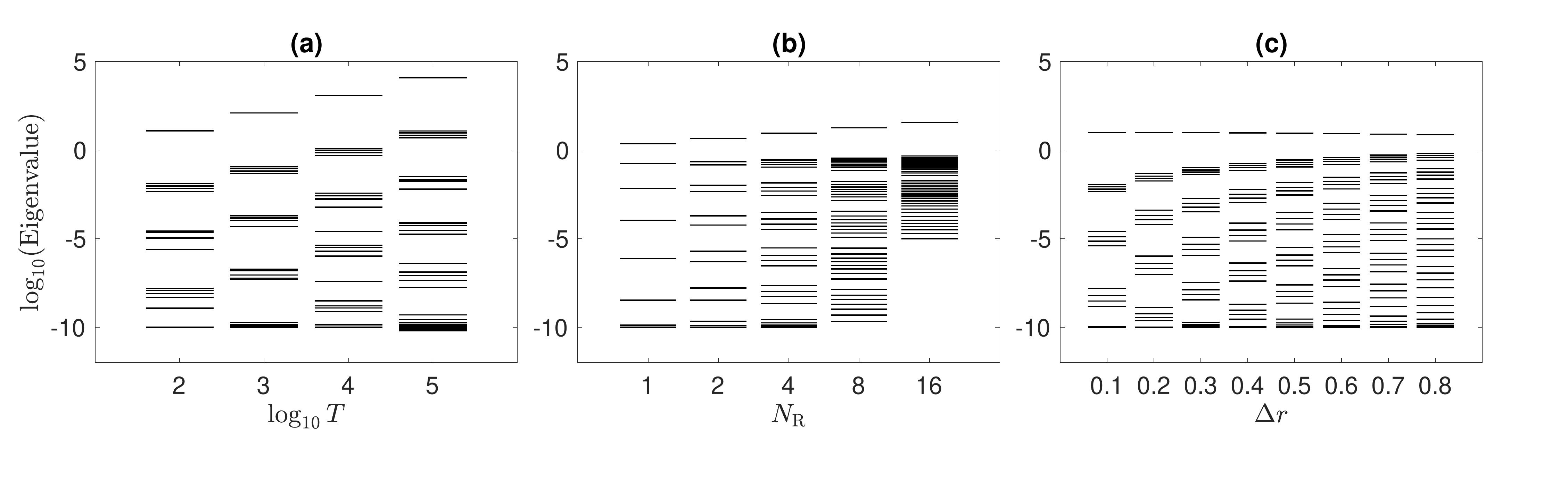}
\caption{(a) Same results as in Fig.~\ref{fig:networkSimulation}, but for a  network of 5 receptors and 40 ligands. All concentrations and binding rates were set to 1. For these simulations, the unbinding rates were chosen uniformly (not log-normally) at random in the range $[0.5, 1.5]$, the rest of parameters are as in Fig.~\ref{fig:networkSimulation}. The eigenvalues group together in subsets of number of receptors ($N_{\rm R} = 5$). These subsets are almost equally spaced on log axis. (b) Eigenvalue spectrum of $\log P''$ vs number of receptors, $T=100$.  The splitting of the eigenvalues changes as the number of receptors increase. (c) Eigenvalue spectrum of $\log P''$ vs the width of the unbinding rate distribution. The unbinding rates were chosen uniformly at random in the range $[1-\Delta r, 1+\Delta r]$, with $N_{\rm R}= 4$, $N_{\rm L}= 40$, $T=100$. Networks with wider ranges of unbinding rates provides better concentration estimates.}
 \end{figure*}

 \begin{figure*}
\centering
\includegraphics[width=0.7\textwidth, trim={0 2.0cm 0 0}]{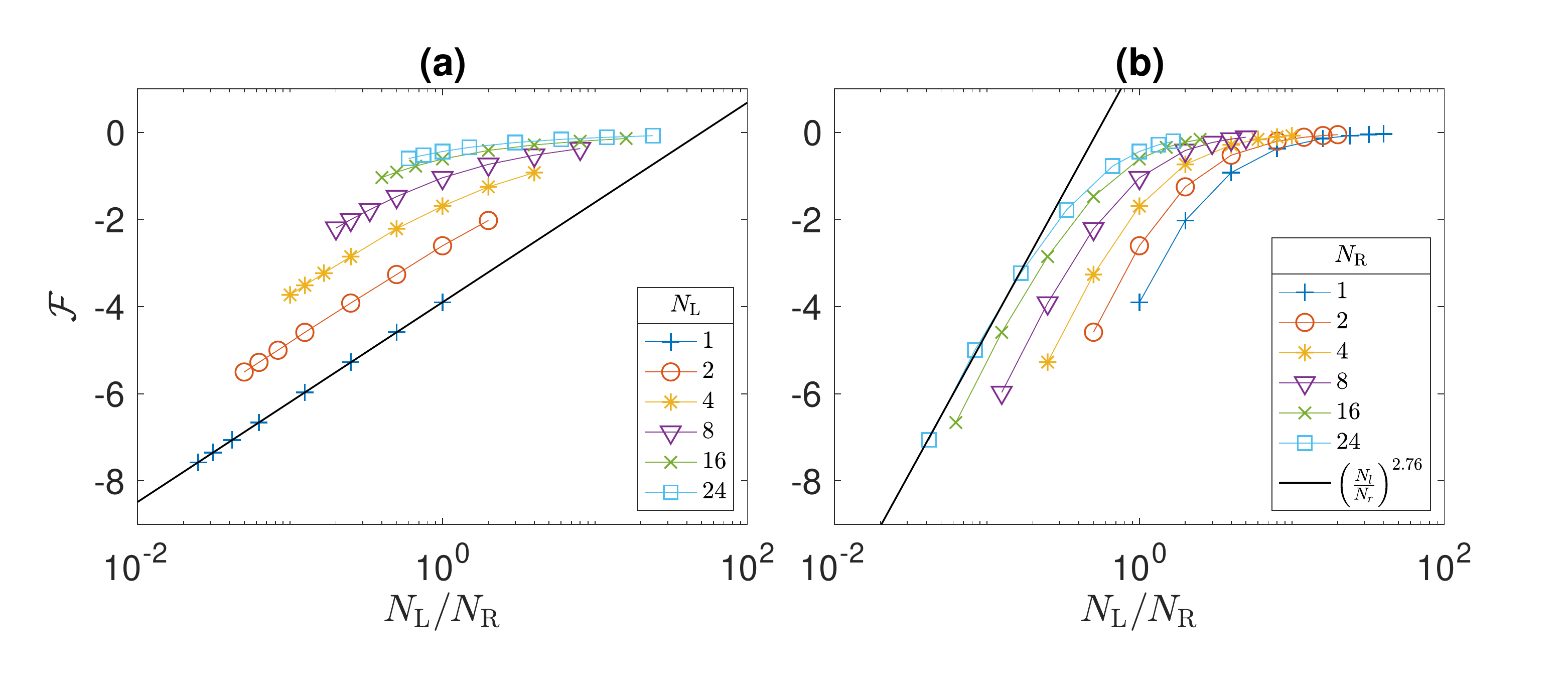}
   \caption{Same analysis as in Fig.~\ref{fig:DetVsL}, but for unbinding rates chosen uniformly (not log-normally) at random in the range $[0.5, 1.5]$. Here $\beta$ is still 1, while $\alpha \approx 2.76$, indicating that, as mentioned before, rates with wider standard deviations make the learning problem easier.}
 \end{figure*}

\end{document}